# Enhancement of electron energy to multi-GeV regime by a dual-stage laser-wakefield accelerator pumped by petawatt laser pulses

Hyung Taek Kim[1,2], Ki Hong Pae[1], Hyuk Jin Cha[1], I Jong Kim[1,2], Tae Jun Yu[1,2], Jae Hee Sung[1,2],

Seong Ku Lee[1,2], Tae Moon Jeong[1,2], and Jongmin Lee[1*]

[1]Advanced Photonics Research Institute, GIST, Gwangju 500-712, Korea

[2]Center for Relativistic Laser Science, Institute for Basic Science GIST-campus, Gwangju 500-712, Korea

leejm@gist.ac.kr*

Laser wakefield acceleration offers the promise of a compact electron accelerator for generating a multi-GeV electron beam using the huge field gradient induced by an intense laser pulse, compared to conventional rf accelerators. However, the energy and quality of the electron beam from the laser wakefield accelerator have been limited by the power of the driving laser pulses and interaction properties in the target medium. Recent progress in laser technology has resulted in the realization of a petawatt (PW) femtosecond laser, which offers new capabilities for research on laser wakefield acceleration. Here, we present a significant increase in laser-driven electron energy to the multi-GeV level by utilizing a 30-fs, 1-PW laser system. In particular, a dual-stage laser wakefield acceleration scheme (injector and accelerator scheme) was applied to boost electron energies to over 3 GeV with a single PW laser pulse. Three-dimensional particle-in-cell simulations corroborate the multi-GeV electron generation from the dual-stage laser wakefield accelerator driven by PW laser pulses.

Since the proposal of laser wakefield acceleration (LWFA) in 1979 [1], the LWFA has been regarded as a novel method for overcoming the limitations of conventional rf-based accelerators. Later, the production of mono-energetic electron beams accelerated in a plasma bubble was theoretically proposed for situations when the pulse length of an intense laser pulse ($a_0 \gg 1$) is shorter than the plasma wavelength ($c \cdot \tau < \lambda_p$) [2]. In 2004, generation of quasi-monoenergetic electron beams with peaks at 70 to 170 MeV was demonstrated in the nonlinear regime using a few tens of TW femtosecond laser systems [3–5]. Afterward, much effort has been pursued to generate stable, tunable, and high-energy electron beams in the 1–2 GeV energy range [6–10]. Recent progress in LWFA and high-power laser technologies has created anticipation of realizing a laser-driven electron accelerator producing energies beyond 100 GeV, which can be used for developing femtosecond x-ray/gamma-ray sources and a laser-based head-on collider [11-13].

Femtosecond high-power laser technology has realized high-performance, repetition-rated, 30-fs, PW-level laser systems for high-field science research [14–16]. Theoretical expectations based on three-dimensional particle-in-cell (PIC) simulations have shown that a PW laser pulse can accelerate electron beams up to multi-GeV energies or even to tens of GeV energies when the injection scheme and the acceleration length are properly chosen [17–19]. Physical parameters, such as the strength of the nonlinear vector potential ($a_0 = eE_0/m_e c\omega_0$), laser depletion length ($L_d \approx (\omega_0/\omega_p)^2 c\tau$), and dephasing length ($L_{dp} \propto \omega_0^2/\omega_p^3$), are crucial factors for increasing electron energy [20]. In general, at a given laser power, lower gas density and longer medium length are desirable for boosting the maximum electron energy by increasing the pump depletion and dephasing lengths. However, in the self-injection scheme, the reduction in gas density can inhibit the electron beam loading to plasma waves and limit the maximum electron energy. A recent report on the scaling of LWFA showed that the gas density for self-injection using 100-TW laser pulses should be over $5 \times 10^{18}$ cm$^{-3}$ [21]. Since the self-injection can limit the energy of an accelerated electron beam in a low-density medium, improving and controlling of the electron beam injection are of crucial importance in generating electron beams at a higher energy.

Several methods to control the injection have been demonstrated by using auxiliary laser pulses [22], manipulating the density gradient [23, 24], applying inner-shell ionization [8, 9], and combining two different gas media [24, 25]. Among those, a simple method is generating electrons in one stage and accelerating them in the other, the so-called staged LWFA or cascaded LWFA [24–28]. Staged LWFA was first demonstrated in the self-modulated regime [26], and it showed tunability of the electron beam energy [24] and a reduction in the electron energy spread [28]. Recently, the generation of a 0.8-GeV monoenergetic electron beam was reported using dual-stage acceleration with two gas cells in the bubble regime [25]. Another advantage of using dual-stage acceleration is that this approach can extend the acceleration length by adding more stages with separate pumping laser pulses, which has been proposed as a method to obtain a 100-GeV electron beam using LWFA driven by multi-beam lines of PW laser pulses [11-13].

In this paper, we report on the multi-GeV electron-beam generation from a dual-stage gas jet pumped by PW laser pulses. Two helium gas jets were used for the dual-stage acceleration; the electron beam was generated from a 4-mm-long gas jet and injected into a 10-mm-long gas jet to boost the electron energy. In the experiment, we observed a substantial increase in the electron energy from the second medium (the 10-mm gas jet) to 3 GeV, which has never been achieved in previous experiments on laser-driven electron acceleration.

The LWFA experiments were performed at the Advanced Photonics Research Institute (APRI) using a PW laser system that can deliver a maximum energy of 30 J with a 30-fs pulse duration. Since the laser chirp affects the stability and beam charge of accelerated electron beams [29, 30], we scanned the laser chirp by detuning the compression grating distance during the experiments and positively-chirped 60-fs pulses (cτ = 18 μm) was chosen for the stability and the maximum electron energy. Figure 1 presents the experimental layout, and the inset shows the scheme of the dual-stage gas jet, consisting of 4-mm- and 10-mm-long gas jets. As shown in Fig. 1, a laser pulse with an energy of 25 J was focused by a spherical mirror, having a focal length of 4 m, onto the dual-stage gas jet. The backing pressures of the two gas jets were independently controlled. The gap between the first and the second jets was about 2 mm, and the laser focus was located in the middle of the gap. The focused beam size was about 25 μm (FWHM) on the target, and the Rayleigh range was about 1.8 mm. The maximum intensity in the focal plane was about $3 \times 10^{19}$ W/cm$^2$, yielding a normalized vector potential of $a_0$ = 3.7. The electron energy distribution was recorded by imaging a Lanex film behind a dipole magnet spectrometer. The electron energy was calibrated by measuring the incident direction of the electron beam on the dipole magnet [31].

We first tested the single-stage electron accelerations using 4-mm and 10-mm jets with PW laser pulses [see Fig. 2(a)]. In this case, the focusing position of the laser beam was located near to the entrance of the gas jet. For the 4-mm gas jet, energetic electrons with a distribution peaking at 400 MeV were obtained at an electron density of $2.1 \times 10^{18}$ cm$^{-3}$. The calculated dephasing length ($L_{dp}$) after trapping was about 7 mm for the 3D nonlinear case [19], and the pump depletion length ($L_p$) was twice the dephasing length. The electrons were accelerated to the end of the 4-mm gas medium, and the maximum electron energy was mainly limited by the acceleration length after the beam loading. For the 10-mm gas jet, electron beams with a distribution peaking at about 800 MeV were obtained at an electron density of $1.1 \times 10^{18}$ cm$^{-3}$. The pump depletion length ($L_p$ = 28 mm) and the dephasing length ($L_{dp}$ = 21 mm) were much longer than the length of the medium that provided the acceleration to the end of the medium. The total charge of each of the electron beams from the 4-mm and 10-mm gas jets was about 88 pC and 110 pC, respectively. The total charge of each of the electron beams from the 4-mm and 10-mm gas jets was about 88 pC and 110 pC, respectively. The charge of the electron beam was measured by an integrating current transformer (ICT) located in between the gas jet target and dipole magnet. The total charge measured by ICT can be different from the charge of a single bunch, and multiple bubbles (or multiple bunches) may contribute to the measured total charge. However, we could not observe an electron signal below 250 MeV to the low limit of the dipole-magnet spectrometer (60 MeV). Thus, the significant part of the total charge can correspond to the charge of the mono-energetic peak except the contributions from low energy electrons below 60 MeV. The divergence of electron beam was 5.5 mrad for 4-mm jet and 5.3 mrad for 10-mm jet.

Figure 2(b) shows the dependence of the peak electron energy on the electron density of the medium. The error bars in the figure represent the standard deviation of 5 laser shots. For the 4-mm gas jet, the peak electron energy was not sensitive to the electron density, as shown in Fig. 2(b). For the 4-mm gas jet, the electron density of the medium was well above the self-injection limit, and it was thought that self-guiding was stable up to the end of this short medium. On the other hand, for the 10-mm gas jet, the electron energy gradually decreased as the electron density increased from the optimal density. Below the electron density of $1.0 \times 10^{18}$ cm$^{-3}$, an energetic electron beam was not generated because self-injection could not occur in this condition due to the low density of the medium.

After the single-stage acceleration, we tested a dual-stage acceleration scheme: the first 4-mm gas jet was used as an injector, and the second 10-mm gas jet, as an accelerator. In this case, the focus of the laser beam was located in the middle of the 2-mm gap between two jets. For the 4-mm jet, the electron generation was not sensitive to the focal position change in few-mm range and we could generate very similar electron beams as shown in Figs. 2(a) and 3(a) with different focal positions for the 4-mm jet. Since the laser intensity was much higher than the critical power for self-focusing with an electron density of $2\times10^{18}$ cm-3, the laser beam could be quickly self-focused in the medium and electron acceleration had been occurred in a similar way in this self-guided plasma with changing the focal position. The 400-MeV electron beams generated from the first gas jet were injected into the 10-mm jet for further acceleration. The highest electron energy of over 3 GeV was observed by lowering the electron density of the second medium to $0.8 \times 10^{18}$ cm$^{-3}$. Note that, at this electron density, we could not generate an electron beam from the single 10-mm gas jet because self-injection did not occur in the medium. In this case, the divergence of the electron beam was about 4 mrad, and the total charge of the beam over the whole spectrum was about 80 pC, while the total charge of the electron beam for energies over 2-GeV is estimated to be about 10 pC. The electron spectrum after the acceleration stage showed a broad energy spread over 1.5 GeV ($\Delta E/E > 50\%$), and two separate peaks were observed at 1.1 GeV and 2.7 GeV while the maximum energy of the electron spectrum reached 4 GeV. The broad energy spread could be partly explained by the low resolution of the electron spectrometer in the high-energy range ($\pm$ 650 MeV at half-maximum).

In order to verify the multi-GeV electron beam generation from the dual-stage acceleration with PW laser pulses, we performed three-dimensional PIC simulations [32, 33] in a boost frame ($\gamma = 8$) by employing the moving window technique. In the reference frame, the total size of a simulation box was set to $0.6 \times 0.6 \times 1.84$ mm3 for the coordinates of (x, y, z) in Figs. 4 (a) and (b), which was resolved by grid cell with a size of ($\lambda$p/8 $\times$ $\lambda$p/8 $\times$ $\lambda$'/32), where $\lambda$' is the laser wavelength in the boost frame and $\lambda$p is the plasma wavelength of injector stage in the lab frame (in total 200 $\times$ 200 $\times$ 4608 grid cells). From PIC simulations, we could observe the increase in electron energy and the energy spread in the electron spectrum. The electron beam generated from the 4-mm injector stage (electron density: $2.2 \times 10^{18}$ cm$^{-3}$) has a peak energy of about 400 MeV and enters the acceleration stage (electron density: $0.9 \times 10^{18}$ cm$^{-3}$), in which the accelerated electron beam reaches energies up to 2.5 GeV [see Fig. 4(c)]. For low electron density, the self-injection is suppressed, and a long dephasing length makes the acceleration uniform to the end of medium. Thus, the PIC simulations support the multi-GeV electron beam generation from a dual-stage (injector and accelerator) acceleration scheme using PW laser pulses.

An injection of electron beams into a plasma bubble is a critical step for the multi-GeV electron generation in the acceleration stage. A larger size of the plasma bubble (radius of bubble, $R \propto$ $1/\sqrt{n_e}$) in the lower-density acceleration stage makes electron injection into the bubble easier. In order to verify the electron injection, we checked the electron density distribution during the interaction. Figures 4(a) and 4(b) show electron density distributions at 3.5 ps and 6.2 ps after the beginning of the interaction, respectively. Here, the electron distributions were shown in the moving window at the moments and the target medium moves from the right to left direction in the simulation window in the reference frame. The laser pulse propagates the full length of the medium during the time from $t = 0$ to $t = 6$ ps in this boost frame. The plasma wave had been created at 3.5 ps by the laser pulse in the acceleration stage, and the electron bunch was loaded into an existing plasma bubble in that stage, as shown in Fig. 4(a) and Fig. 4(a′). At the end of the interaction ($t = 6.2$ ps), the injected electrons (the seed beam) are accelerated through the remaining length of the medium [Figure 4(b)]. Finally, as shown in Fig. 4(c), the electron beams generated in the injection stage attain an energy of 2.5 GeV in the acceleration stage.

Because the combination of medium densities is a critical parameter in determining the electron energy, we examined the dependence of the electron energy on the density matrix between the two gas jets. Figure 5 shows the electron energy map obtained with different electron densities for the two gas jets. The map shows two different regimes for the electron energy with respect to the electron densities of the two jets. In one regime (with higher density for the injector and lower density for the accelerator), the second medium acts as an accelerator for boosting electron energy. As the electron density of the accelerator stage increases, the maximum electron energy decreases because the electron bunch can be injected into the deceleration phase of the second bubble due to the smaller bubble size. In the other acceleration regime, the two gas media have similar electron densities of about $2.2 \times 10^{18}$ cm$^{-3}$. In this case, the second medium acts as an extended medium with a density dip in the gap, and the peak electron energy is about 1.1 GeV.

In summary, we observed a significant improvement in electron energy to 3 GeV by using a dual-stage acceleration scheme with PW laser pulses. The physical properties of the electron beams, such as beam charge, divergence, and spatial beam profile, were investigated. Three-dimensional PIC simulation in a boost frame supported the observed multi-GeV electron generation from the dual-stage accelerator. By investigating the electron energy map with a two-dimensional-density matrix, we found that two kinds of acceleration regimes were possible in the dual-stage acceleration scheme. This result can be considered as an important step forward in the development of a compact multi-GeV electron accelerator, achieved by optimizing a dual-stage accelerator with PW laser pulses; such an accelerator can be applied to generate femtosecond high-flux gamma-ray sources and coherent hard x-ray sources. According to the formula by W. Lu et al. [19], we can estimate the optimized parameters for achieving 10-GeV electron beams with dual-stage accelerator. A 10-GeV energy gain in LWFA can be achievable with the medium density of $1 \times 10^{17}$ cm-3, medium length of 45 cm, pulse duration of 80 fs and beam spot size of 60 μm, when we assume the energy depletion of 50% in the first medium out of an initial pumping energy of 30 J. Under this condition, the external-guiding technique is necessary because the laser power in the second medium is lower than the critical power for self-guiding. In addition, the dual-stage acceleration can be used to show the feasibility of multi-stage acceleration with proper laser and medium conditions, and it might provide a pathway to realize a 100-GeV electron beam in the near future.

This work was supported by the Ministry of Knowledge and Economy of Korea through the Ultrashort Quantum Beam Facility Program. This work was also supported by the Research Center Program of IBS (Institute for Basic Science) in Korea and Applications of Femto-Science to Nano/bio-Technology Utilizing Ultrashort Quantum Beam Facility through a grant provided by GIST.

# Figure Captions

**FIG. 1.** (color). Experimental layout. The dipole magnet has length of 30 cm and magnetic field strength of 1.33 T, which was installed 1 m away from the gas-jet target. Two Lanex screen have been installed at the entrance and exit of the magnet to measure electron beam profile and energy, respectively. The ICT was installed between gas jet and dipole magnet to measure the charge of the electron beam.

**FIG. 2.** (color). (a) Electron energy spectrum for 10-mm [red line and image (i)] and 4-mm [blue line and image (ii)] gas jets. (b) Electron energy with respect to the electron density.

**FIG. 3.** (color). (a) Electron energy spectrum and (b) spatial profile for the seed beam from the 4-mm target [red line and (i)] and the accelerated electrons from the dual-stage target [black line and (ii)].

**FIG. 4.** (color). Electron density distribution at (a) $t = 3.5$ ps and (b) $t = 6.2$ ps and (c) the electron energy spectrum of the seed beam and accelerated beam in the dual-stage accelerator calculated by 3D PIC simulation in the boost frame ($\gamma = 8$).

**FIG. 5.** (color). Electron energy map as a function of electron densities of the two gas jet media.

**Figure 1**

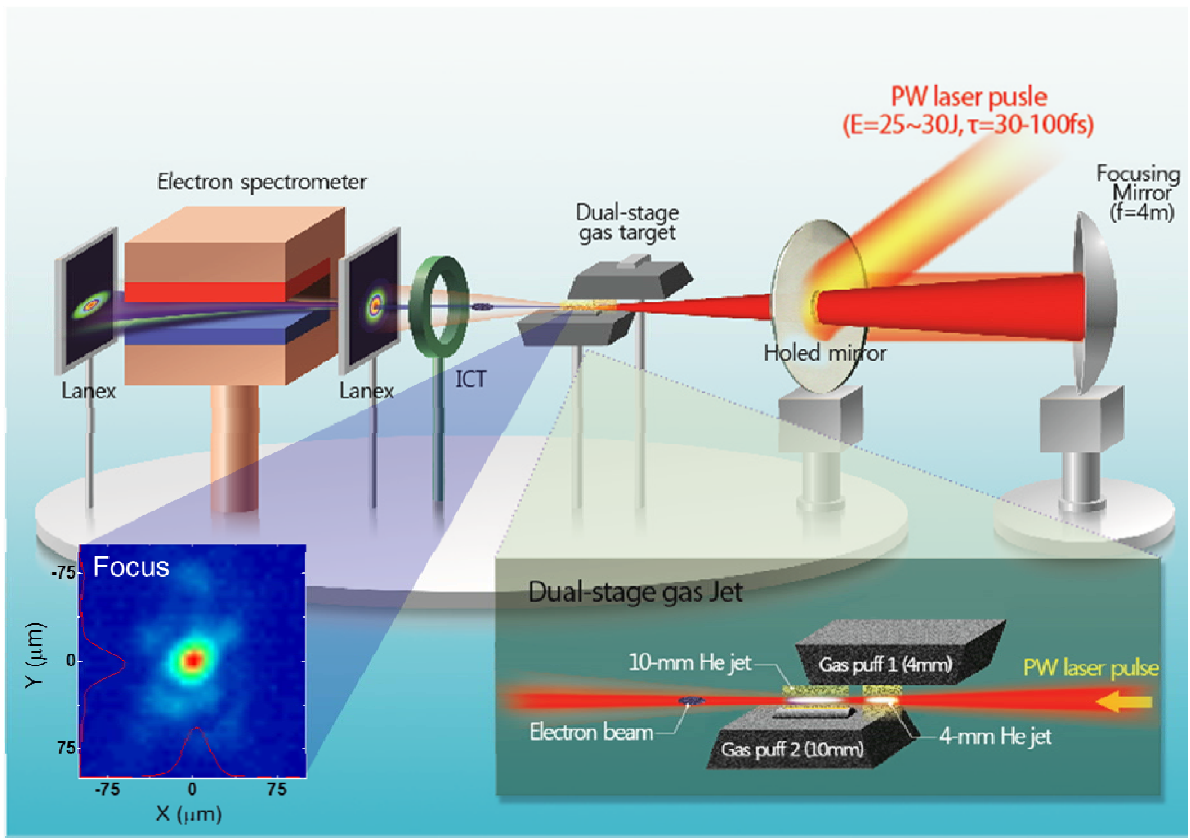

**Figure 2**

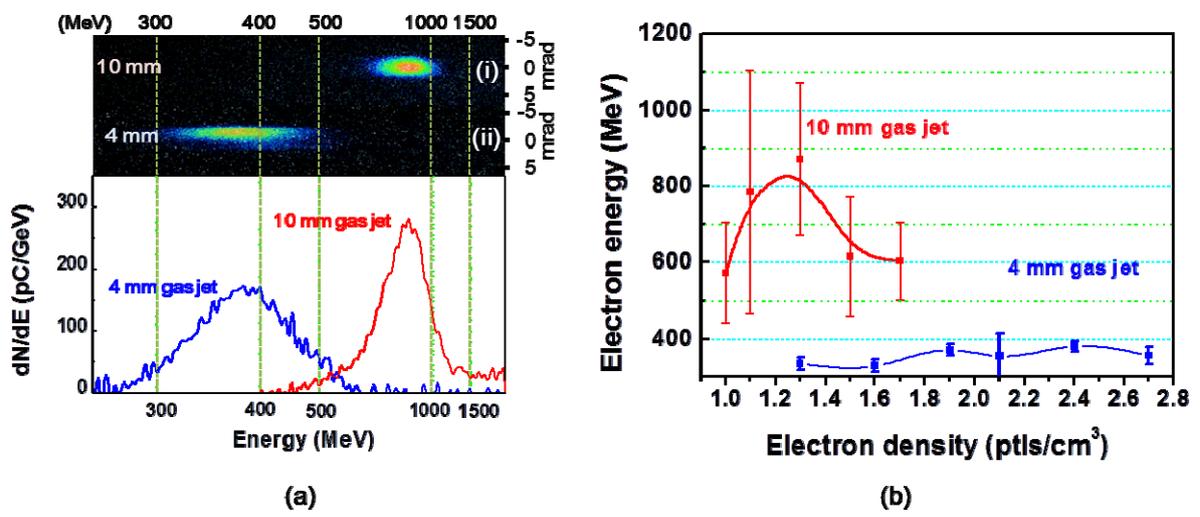

**Figure 3**

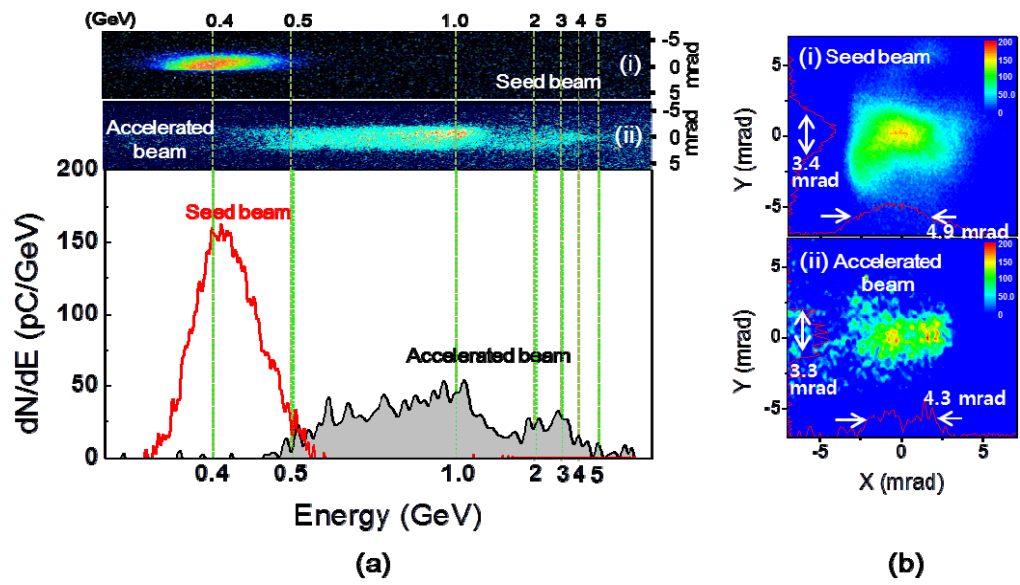

(a) (b)

**Figure 4**

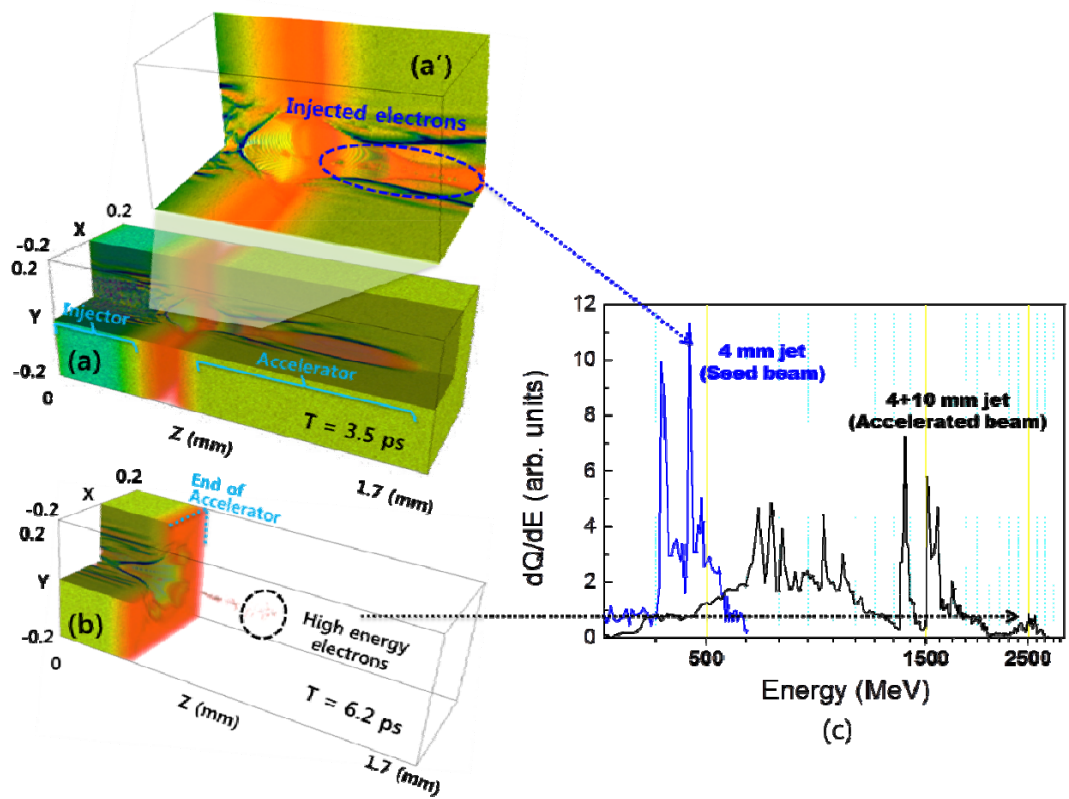

**Figure 5**

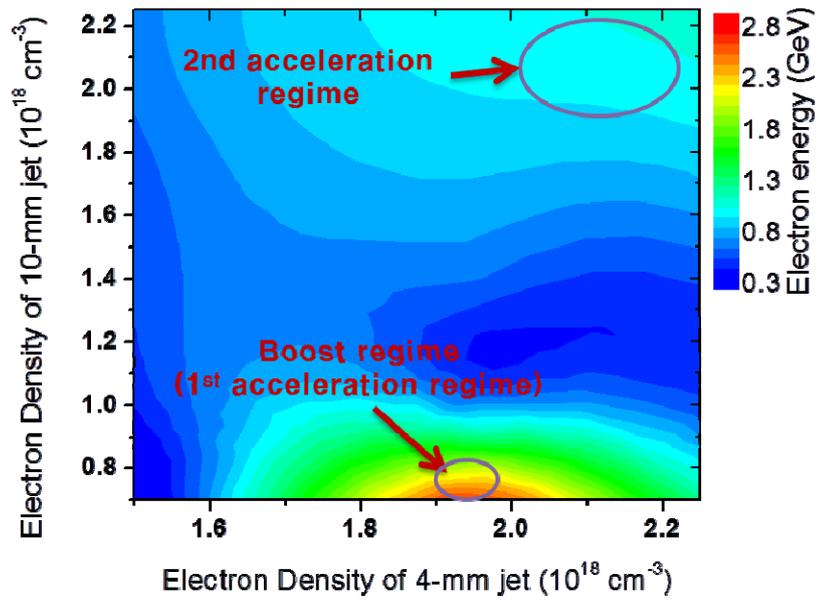

# Supplemental material for

# Multi-GeV electron beam generation from a dual-stage laser-wakefield accelerator driven by PW laser pulses

Hyung Taek Kim[1,2], Ki Hong Pae[1], Hyuk Jin Cha[1], I Jong Kim[1,2], Tae Jun Yu[1,2], Jae Hee Sung[1,2],

Seong Ku Lee[1,2], Tae Moon Jeong[1,2], and Jongmin Lee[1*]

1Advanced Photonics Research Institute, GIST, Gwangju 500-712, Korea

2Center for Relativistic Laser Science, Institute for Basic Science GIST-campus, Gwangju 500-712, Korea

leejm@gist.ac.kr*

## 1. PW Laser system

The laser-wakefield-electron-acceleration experiment has been performed using a petawatt Ti:sapphire laser system operating at a 0.1-Hz repetition rate in Advanced Photonics Research Institute, GIST, named as PULSER (PW Ultrashort Laser System for Extreme Science Research). The laser system based on a chirped-pulse amplification technique is composed of a 1-kHz multipass-amplifier front-end system, a grating stretcher, two power amplifiers, three-pass booster amplifier, and a grating compressor. The laser pulse is amplified to a maximum energy of 47 J from the amplifiers. After passing through the grating compressor, the laser pulse has 34-J energy and 30-fs duration, resulting in 1.1-PW peak power [1]. This 0.1-Hz 1.1-PW laser pulse, which has an amplified spontaneous emission level of $10^{-8}$ at 100 ps before the main pulse, was focused onto the gas target. A deformable mirror was installed before the compressor for the pre-compensation of wave-front aberration measured by wavefront sensors.

## 2. Gas Jet System

The dual-stage He-gas jet consisted of 4-mm- and 10-mm-long gas nozzles that have a width of 1.2-mm [2, 3]. The gas densities from the nozzles were measured by Mach-Zehnder interferometry [4] at different backing pressures. The backing pressure and nozzle height were independently controlled in the range of 0.5~5 bar and 1 ~ 4 mm, respectively. The first 4-mm-long gas nozzle was installed to puff downward while the second 10-mm-long one was puffing upward as shown in the inset of Fig 2. The gap between two He-jets was set to 2 mm after optimizing the gap distance by scanning the gap from 0 to 6 mm.

## 3. Electron Spectrometer

The electron spectra were measured by a dipole-magnet-electron spectrometer with B=1.33 T and 30-cm length. In order to avoid calibration error induced by the pointing instability of electron beam, we simultaneously measured the electron beam pointing in front of the dipole magnet, and the electron energy was recalculated with the corresponding direction of electron beam to the dipole magnet. The method of calibrating the electron energy was described in Ref [5].

## 4. PIC simulations

Three-dimensional particle-in-cell simulations were carried out using the fully relativistic electromagnetic code ALPS [6-8] with the boost-frame technique [9]. The laser pulse was modeled by sin2 temporal (60 fs FWHM duration) and Gaussian spatial profile. The targets were modeled by cold plasmas with electrons and immobile background ions. The electron density profile of the target plasma was modeled using the experimentally measured data. The longitudinal mesh size was set to λ0'/32 (where λ0' is the laser wavelength in the boost frame) and the transverse mesh size was chosen to minimize the numerical dispersion error. In all simulations, fields at a given particle position was determined using cubic interpolation to push macro-particles, and an exact charge conservation scheme [10] was used for the current density deposition. The electron energy spectrum was calculated from the particle data by taking a simple Lorentz transformation after a sufficient propagation in vacuum. We tested the boost γ-factor of 14 and 16 and the results showed consistent convergence. To suppress the numerical noise, we used a compensation binomial filter similar to that described in Ref. [11].